\documentclass[twocolumn]{jjap2} 
\usepackage{bm}
\title{Synchronized Magnetization Oscillations in F\slash  N\slash F Nanopillars}
\author{
Kiwamu \textsc{Kudo}\thanks{kiwamu.kudo@toshiba.co.jp}, 
Rie \textsc{Sato} 
and Koichi \textsc{Mizushima}
}
\inst{Corporate Research and Development Center, Toshiba Corporation, 
Kawasaki, 212-8582, Japan}
\abst{
Current-induced magnetization dynamics in a trilayer structure 
composed of two ferromagnetic free layers and a nonmagnetic spacer is examined.  
Both free layers are treated as a monodomain 
magnetic body with an  uniform magnetization. 
The dynamics of the two magnetizations is modeled by  
modified Landau-Lifshitz-Gilbert equations with spin-transfer torque terms. 
By solving the equations simultaneously, 
we discuss their various solutions in detail. 
We show that there exists the synchronous motion of two magnetizations 
among the various solutions; 
the magnetizations are resonantly coupled via spin-transfer torques and perform  
precessional motions with the same period. 
The condition to excite the synchronous motion depends on 
the difference between the intrinsic frequencies of the two ferromagnetic free layers 
as well as the magnitude of current. 
}
\kword{spin-transfer torque, current-induced magnetization dynamics, 
Landau-Lifshitz-Gilbert equation, 
synchronization, magnetization oscillations}
\begin{document}
\maketitle
\section{Introduction}
Current-induced magnetization excitations in magnetic nanopillars, 
predicted by Slonczewski \cite{Slonczewski} and  Berger \cite{Berger} in 1996, 
have been extensively studied. 
The basic structure of nanopillars is 
ferromagnetic\slash nonmagnetic\slash ferromagnetic (F\slash N\slash F) 
trilayer structure. 
It has been demonstrated experimentally that 
a spin-polarized dc current through a 
``free'' ferromagnetic layer can reverse 
its magnetization \cite{Tsoi,Myers,Katine,Grollier,Sun,Urazhdin}. 
Moreover, recent experiments have shown that 
a coherent precession of the magnetization at GHz frequencies 
is induced by a spin-polarized current 
and that its frequency of precession depends on 
applied field and current density \cite{Kiselev,Rippard,Krivorotov}. 
These magnetization dynamics, 
current-induced  magnetization switching (CIMS) and 
coherent precession, 
provides the possibility 
to utilize the spin-transfer phenomena 
for various applications, 
such as magnetic random access memory cells, 
nanometer-sized microwave generators, and so on.

Most experimental studies about CIMS and microwave generation have concerned 
an ``asymmetric'' trilayer 
consisting of a free thin ferromagnetic layer 
and a thick ferromagnetic layer with a fixed magnetization. 
Accordingly, many theoretical works 
also have concerned 
the asymmetric structure \cite{Sun2,Li,Bazaliy,Bertotti,Xiao,Gorley}. 
By active experimental and theoretical studies for the past ten years, 
many properties of magnetization dynamics 
in the asymmetric structure have already been clarified. 
On the other hand, the  magnetization dynamics in a 
``symmetric'' structure 
where both ferromagnetic layers play the same role as free layers 
has received little attention. 
It is reported recently that 
precessional dynamics of the thick layer as well as 
the thin layer are excited by current in the usual asymmetric sample \cite{Kiselev2}. 
Therefore, 
it is important to examine 
what kind of magnetization dynamics can be excited in the 
two-free-layers structures.

\begin{figure}
\begin{center}
\includegraphics[width=40mm]{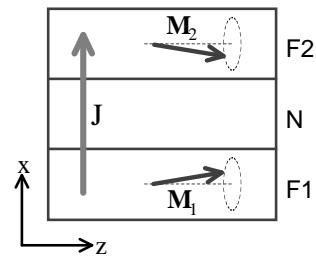}
\end{center}
\caption{Schematic of the F1/N/F2  trilayer. 
         F1 and F2 are thin ferromagnetic ``free'' layers. 
         N is a  nonmagnetic spacer. 
         $J$ is current density perpendicular to plane. 
         F1 and F2 are treated as a monodomain ferromagnet. }
\label{fig:1}
\end{figure}

In this paper, we examine current-induced magnetization dynamics 
in the F1/N/F2 nanopillars where 
both F1 and F2 are ferromagnetic free layers and N is a nonmagnetic spacer 
as illustrated in Fig.~\ref{fig:1}. 
By the current passing through the trilayer, 
the magnetizations of F1 and F2, $\bm{M}_1$ and $\bm{M}_2$, interact 
with each other via spin-transfer torques \cite{Slonczewski}, 
and perform various motions. 
We discuss these motions in detail comparing with the magnetization dynamics 
in an asymmetric structure. 
We show that among these various motions there exists 
a synchronous motion of two magnetizations. 
In the motion, the two magnetizations perform a stable precessional motion 
with the same period, 
analogous to the synchronized oscillation of two coupled 
nonlinear oscillators. 
The synchronization of magnetizations may be efficient for raising 
the power of microwave radiation emitted by 
the magnetic multilayers which function as microwave generators.

\section{Model}
With regard to two ferromagnetic layers in the F1\slash N\slash F2 trilayer, 
we assume that both F1 and F2 have an uniaxial anisotropy in the $z$ direction. 
We assume further that F1 and F2 have almost identical properties 
except that there is a difference between the magnitude of the effective magnetic fields 
acting on $\bm{M}_1$ and $\bm{M}_2$. 
Within monodomain approximation, we 
describe the dynamics of the two magnetization vectors, 
$\bm{M}_1$ and $\bm{M}_2$, by the modified Landau-Lifshitz-Gilbert equations
\begin{equation}
\frac{d \bm{m}_i}{d   \tau}
=-\bm{m}_i \times \bm{h}_\mathrm{eff}^{(i)}
+\alpha \bm{m}_i \times \frac{d \bm{m}_i}{d   \tau}
+\bm{\Gamma}_i,
\label{eq:LLG}
\end{equation}
with the spin-transfer torque term $\bm{\Gamma}_i$. 
Here we have used the dimensionless form of LLG equations. 
In eqs.~(\ref{eq:LLG}), 
$\bm{m}_i=\bm{M}_i\slash M_\mathrm{s}$, $M_\mathrm{s}$ is 
the saturation magnetization, 
times $\tau$ is measured in units of $(\gamma H_\mathrm{u}^{(1)})^{-1}$, 
$H_\mathrm{u}^{(1)}$ is the magnitude of 
the uniaxial anisotropy field acting on $\bm{M}_1$, 
$\gamma$ is the gyromagnetic ratio, 
and $\alpha$ is the dimensionless Gilbert damping constant of which 
we always take $0.01$ in this paper. 
$\bm{h}_\mathrm{eff}^{(i)}$ is the effective magnetic field 
normalized by $H_\mathrm{u}^{(1)}$ and given by 
\begin{equation}
\bm{h}_\mathrm{eff}^{(i)}
=(h_\mathrm{ext}^{(i)}+h_\mathrm{u}^{(i)}m_{iz})\hat{\bm{z}}
-h_\mathrm{p}^{(i)}m_{ix}\hat{\bm{x}},
\label{eq:eff_field}
\end{equation}
where 
$h_\mathrm{ext}^{(i)}$ is the external field applied along the $z$ direction, 
$h_\mathrm{u}^{(i)}$ is the in-plane uniaxial-anisotropy field 
($h_\mathrm{u}^{(1)}=1.0$), 
and $h_\mathrm{p}^{(i)}(\simeq 4 \pi M_\mathrm{s}\slash H_\mathrm{u}^{(1)})$ is the 
effective out-of-plane anisotropy field due to the film shape of ferromagnets. 

The spin-transfer torque term $\bm{\Gamma}_i$ 
acting on $\bm{M}_i$ can be written in the form \cite{Slonczewski,Li} 
\begin{equation}
\bm{\Gamma}_i
=-a_J \bm{m}_i \times (\bm{m}_1 \times \bm{m}_2). 
\label{eq:STT}
\end{equation}
The parameter $a_J$ represents the strength of the spin-transfer torque and is 
proportional to current density $J$. 
We assign the positive value of $a_J$ for the current flowing from F1 to F2. 
It is noticed that the strength of 
the spin-transfer torque acting on $\bm{m}_1$ and $\bm{m}_2$ are 
equal because we have assumed that F1 and F2 have almost identical properties. 
By means of eqs.~(\ref{eq:STT}), the two magnetizations can interact with 
each other. 
The motion of $\bm{m}_2$ contributes 
to that of $\bm{m}_1$ via $\bm{\Gamma}_1$, and vice versa. 
Due to the form of eqs.~(\ref{eq:STT}), 
$\bm{\Gamma}_i=\bm{0}$ when $\bm{m}_1$ and $\bm{m}_2$ are parallel. 

We solve eqs.~(\ref{eq:LLG}) numerically 
and examine the current-induced dynamics of $\bm{m}_i$ 
on the unit sphere $|\bm{m}_i|=1$ \cite{note}. 
We conduct the calculation using mainly the two parameter sets for the effective fields; 
(i) $h_\mathrm{ext}^{(1)}=h_\mathrm{ext}^{(2)}=2.0$, 
$h_\mathrm{u}^{(1)}=1.0$, $h_\mathrm{u}^{(2)}=1.0+\Delta h_\mathrm{u}$, 
$h_\mathrm{p}^{(1)}=h_\mathrm{p}^{(2)}=0.0$, and 
(ii) $h_\mathrm{ext}^{(1)}=1.0$, 
$h_\mathrm{ext}^{(2)}=1.0+\Delta h_\mathrm{ext}$, 
$h_\mathrm{u}^{(1)}=h_\mathrm{u}^{(2)}=1.0$, 
$h_\mathrm{p}^{(1)}=h_\mathrm{p}^{(2)}=10.0$. 
In the parameter set (i), we neglect the effect of 
out-of-plane demagnetizing fields $h_\mathrm{p}^{(i)}$ for simplicity. 
On the other hand, the parameter set (ii) is realistic and experimentally realizable. 
In the both sets of parameters, 
we model the difference between the magnitude of effective magnetic fields 
acting on the two magnetizations by 
$\Delta h_\mathrm{u}$ and $\Delta h_\mathrm{ext}$, respectively. 
We assume that $\Delta h_\mathrm{u}\ge 0$ and $\Delta h_\mathrm{ext}\ge 0$. 
Since $\Delta h_\mathrm{u}\ge 0$ or $\Delta h_\mathrm{ext}\ge 0$, 
we consider the trilayers where the magnitude of effective magnetic field 
acting on $\bm{m}_2$ is larger than that acting on $\bm{m}_1$ 
in both models (i) and (ii). 
As we discuss below, 
the difference of the effective magnetic fields, 
such as $\Delta h_\mathrm{u}$ and $\Delta h_\mathrm{ext}$, 
has the essential role to arise a synchronous motion of 
the two magnetizations. 
By using the simpler model (i), 
we clarify the characteristics of the synchronous motion of the magnetizations. 
By using the model (ii), 
we show that the synchronous motion exists even in a realistic setup.  
The parameters in (ii) cover the essential features of magnetization dynamics 
in the almost symmetric F\slash N\slash F trilayers including 
the two free layers 
with $h_\mathrm{u}^{(i)} \leq h_\mathrm{ext}^{(i)} \ll h_\mathrm{p}^{(i)}$. 
Note that $h_\mathrm{p}^{(i)}$ is several tens of times larger than 
$h_\mathrm{u}^{(i)}$ in usual film ferromagnets. 
Moreover, $\Delta h_\mathrm{ext}$ can be realized 
by a exchange-bias field as that in spin-valve pillars.

In both models (i) and (ii), the equilibrium direction of the two magnetizations 
along $\bm{h}_\mathrm{eff}^{(i)}$ corresponds to $\bm{m}_i=\hat{\bm{z}}$ 
in the absence of current. 
Accordingly, it would be appropriate that $\bm{m}_i=\hat{\bm{z}}$ 
as the initial value at $\tau=0$. 
However, when $\bm{m}_1$ and $\bm{m}_2$ are completely parallel at $\tau=0$, 
the spin-transfer torque, eq.~(\ref{eq:STT}), does not work at all and 
any magnetization dynamics are not excited. 
Therefore, 
we permit that the two magnetizations shift slightly 
from their equilibrium directions and are initially not parallel each other. 
This is justified by taking into account the effect of finite temperature. 
In the real materials with temperature $T \neq 0$, 
thermal fluctuations always exist and then 
magnetizations fluctuate around their equilibrium points 
all the time. 
Therefore, we assume that two magnetizations are initially 
in the vicinity of their equilibrium points and are not parallel, 
although we do not take account of thermal fluctuations 
when we solve eqs.~(\ref{eq:LLG}).

\section{Synchronized Precession of Two Magnetizations}
In this section, 
we discuss the synchronous motion of the magnetizations by 
showing the results of the calculation in model (i), i.e., 
$\alpha=0.01$, $h_\mathrm{ext}^{(1)}=h_\mathrm{ext}^{(2)}=2.0$, 
$h_\mathrm{u}^{(1)}=1.0$, $h_\mathrm{u}^{(2)}=1.0+\Delta h_\mathrm{u}$, 
$h_\mathrm{p}^{(1)}=h_\mathrm{p}^{(2)}=0.0$. 
We consider only the case 
that the applied current is positive, $a_J>0$, 
because the synchronization of two magnetizations do not occur when $a_J<0$ 
in model (i). 

\subsection{Phase diagram}
\begin{figure}
\begin{center}
\includegraphics[width=48mm]{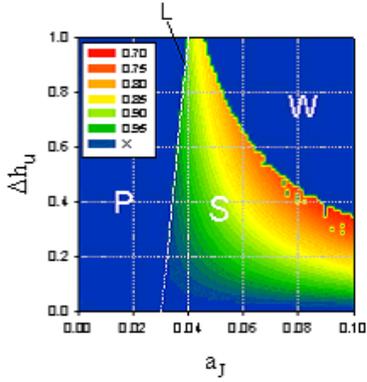}
\end{center}
\caption{(Color) Dynamical stability diagram for $a_J$ and $\Delta h_\mathrm{u}$. 
S denotes synchronized precession of two magnetization, 
P parallel configuration, 
and W chaotic motions. 
The numerical values in the legend represent the steady-state values of $m_{1z}$. 
X represents parallel configuration or chaotic motions. 
The broken line L represents the threshold for current-driven excitations. }
\label{fig:2}
\end{figure}

Depending on $a_J$ and $\Delta h_\mathrm{u}$, 
several distinct types of dynamical modes as the solutions of eqs.~(\ref{eq:LLG}) 
are excited by the spin-transfer torque. 
Figure~\ref{fig:2} shows the dependence of steady state solutions 
on $a_J$ and $\Delta h_\mathrm{u}$. 
In the region labeled S, the synchronized precession of two magnetizations 
on which we will focus in this section occurs. 
In the other regions which are labeled by P and W, 
any kind of coherent precessions do not exists. 
The region P denotes the region 
where a static parallel configuration of two magnetization occurs. 
The region W denotes the region 
where two magnetizations behave chaotic.

\begin{figure}
\begin{center}
\includegraphics[width=40mm]{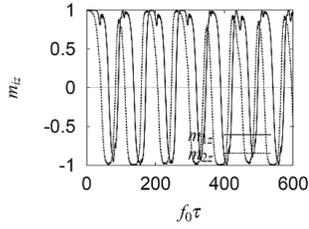}
\end{center}
\caption{Typical behaviors of $\bm{m}_1$ and $\bm{m}_2$ 
in the region W. 
Time evolution of $m_{iz}(\tau)$ is shown. 
The parameters are chosen as $a_J=0.08$ and $\Delta h_\mathrm{u}=0.8$. 
The initial conditions are chosen as $m_{1z}(0)=0.998$, $m_{2z}(0)=0.999$, and 
$\varphi=\pi \slash 6$.}
\label{fig:3}
\end{figure}

Figure~\ref{fig:3} shows the typical behaviors of $\bm{m}_1$ and $\bm{m}_2$ 
in the region W. 
It is found that the behavior like magnetization switching arises irregularly. 
To obtain Fig.~\ref{fig:3}, 
we have used $m_{1z}(0)=0.998$, $m_{2z}(0)=0.999$, and $\varphi(0)=\pi\slash 6$ 
as initial conditions. 
Here, $\varphi=\varphi_2-\varphi_1$ 
where $\varphi_i$ is the azimuthal angle of $\bm{m}_i$. 
In other words, $\varphi$ is the angle between $\bm{m}_1^\perp$ and 
$\bm{m}_2^\perp$ which are the projections of $\bm{m}_1$ and $\bm{m}_2$ 
onto the $xy$-plane, respectively; see Fig.~\ref{fig:5}(d). 
In the horizontal axis of Fig.~\ref{fig:3}, 
$f_0$ is the intrinsic frequency of $\bm{m}_1$ 
in the absence of current. 
$f_0=\omega_0\slash 2\pi$, 
where $\omega_0$ is given by 
the well-known formula, 
\begin{equation}
\omega_0=\sqrt{(h_\mathrm{ext}^{(1)}+h_\mathrm{u}^{(1)})
(h_\mathrm{ext}^{(1)}+h_\mathrm{u}^{(1)}+h_\mathrm{p}^{(1)})}. 
\label{eq:w_0}
\end{equation}
$\omega_0$ can be determined by the ferromagnetic resonance (FMR) experiment. 
We have introduced $f_0$ as the standard precessional frequency. 
In model (i), $f_0=3.0\slash 2 \pi$.

\begin{figure}
\begin{center}
\includegraphics[width=30mm]{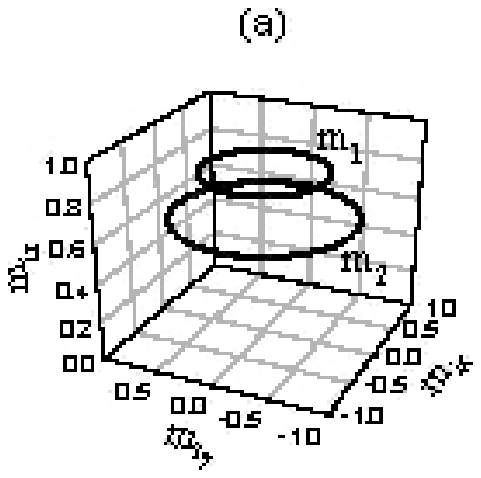}\\
\includegraphics[width=37mm]{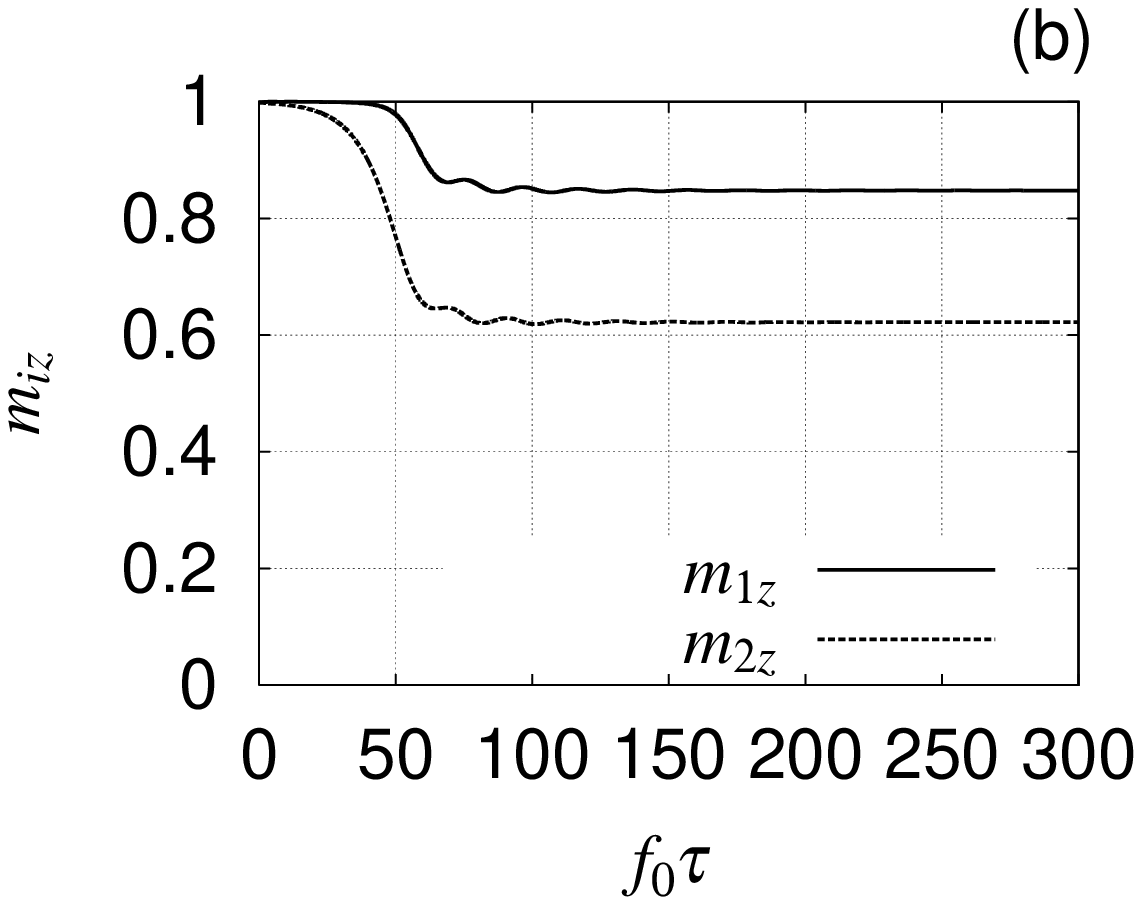}
\includegraphics[width=37mm]{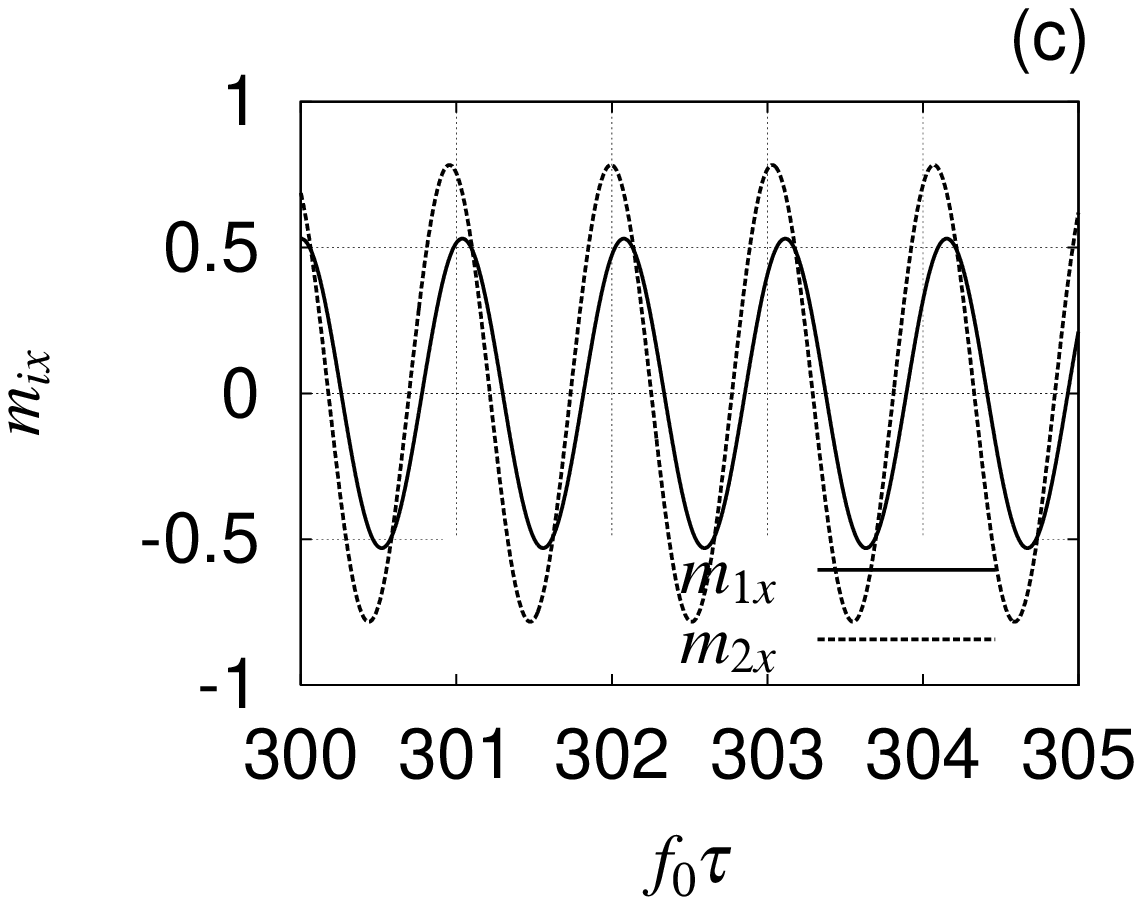}
\end{center}
\caption{Typical behaviors of $\bm{m}_1$ and $\bm{m}_2$ 
in the region S. The parameters are chosen as $a_J=0.06$ and $\Delta h_\mathrm{u}=0.4$. 
The initial conditions are chosen as $m_{1z}(0)=0.998$, $m_{2z}(0)=0.999$, and 
$\varphi=\pi \slash 6$. 
(a) Bird's-eye view of trajectories in the steady state. 
(b) Time evolution of $m_{iz}(\tau)$. 
(c) Time evolution of $m_{ix}(\tau)$ at intervals between $f_0 \tau=300$ and $305$.}
\label{fig:4}
\end{figure}

In the region S, 
both $\bm{m}_1$ and $\bm{m}_2$ precess around the $z$ axis with 
circular forms. 
The circular trajectories of the precessional motion of 
$\bm{m}_1$ and $\bm{m}_2$ are shown in Fig.~\ref{fig:4}(a). 
The steady precessions occurs after transient behaviors. 
The behaviors are shown in Fig.~\ref{fig:4}(b) 
where 
the time evolutions of $m_{1z}(\tau)$ and $m_{2z}(\tau)$ are plotted. 
Figure.~\ref{fig:4}(c) shows 
the time evolutions of the $x$ components, $m_{1x}(\tau)$ and $m_{2x}(\tau)$, 
in the steady state at intervals between $f_0 \tau=300$ and $305$. 
As can be recognized from Fig.~\ref{fig:4}(c), 
the precessional frequencies of $\bm{m}_1$ and $\bm{m}_2$ are equal. 
In other words, the precessional phase is mutually locked. 
That is why we name the magnetization dynamics in the region S 
``synchronized precession'' of two magnetizations.

\subsection{Synchronized precessions}
We discuss the properties of the synchronized precessions in detail. 

The precessional amplitudes become large as $a_J$ and $\Delta h_\mathrm{u}$ are 
increased. 
The dependence of the amplitude of $\bm{m}_1$ 
on $a_J$ and $\Delta h_\mathrm{u}$ has been shown in Fig.~\ref{fig:2}. 
The legend in Fig.~\ref{fig:2} represents 
the values of $m_{1z}$ in the synchronized precessional state. 
Although it is not illustrated, 
the amplitude of $\bm{m}_2$ has the similar dependence on 
$a_J$ and $\Delta h_\mathrm{u}$ as that of $\bm{m}_1$. 
The difference between the precessional amplitude of $\bm{m}_1$ and 
$\bm{m}_2$ is that 
the amplitude of $\bm{m}_2$ is larger than 
that of $\bm{m}_1$ as is shown in Fig.~\ref{fig:4}(a) 
for any $a_J$ and $\Delta h_\mathrm{u}$. 
This property that $m_{1z}>m_{2z}$ is essential for the stability 
of the synchronous precession. 

\begin{figure}
\begin{center}
\includegraphics[width=35mm]{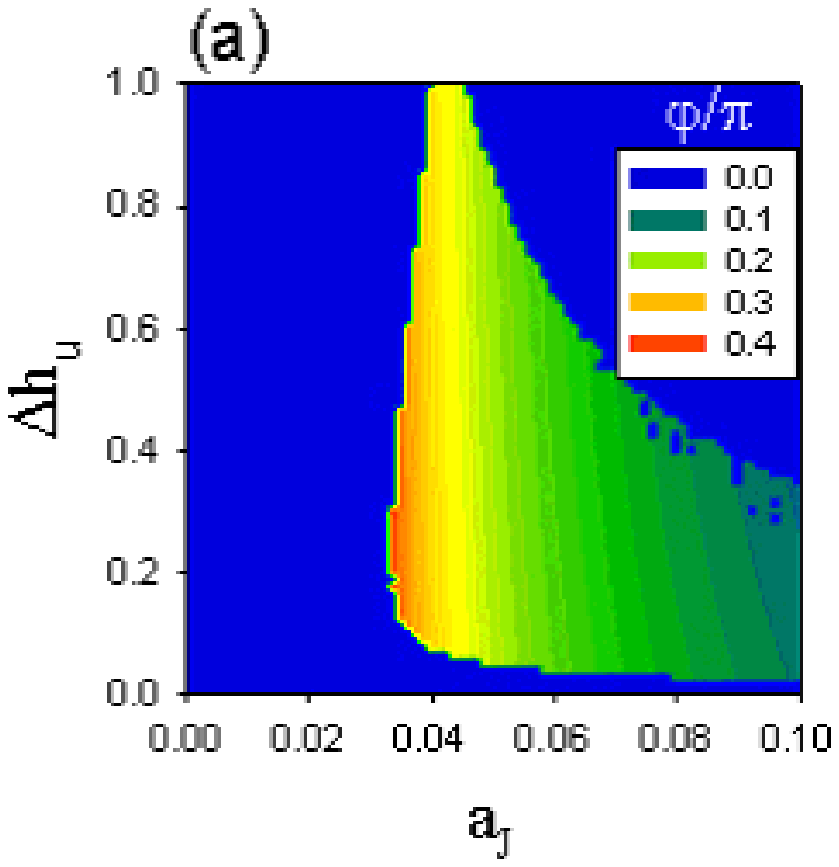}
\includegraphics[width=35mm]{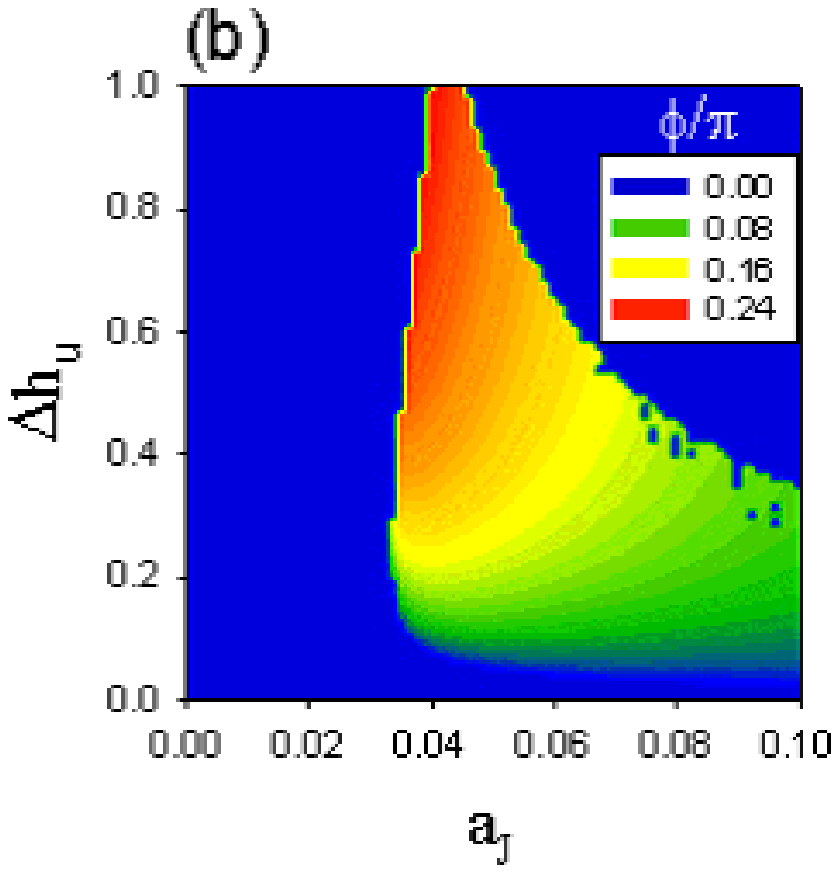} \\
\includegraphics[width=35mm]{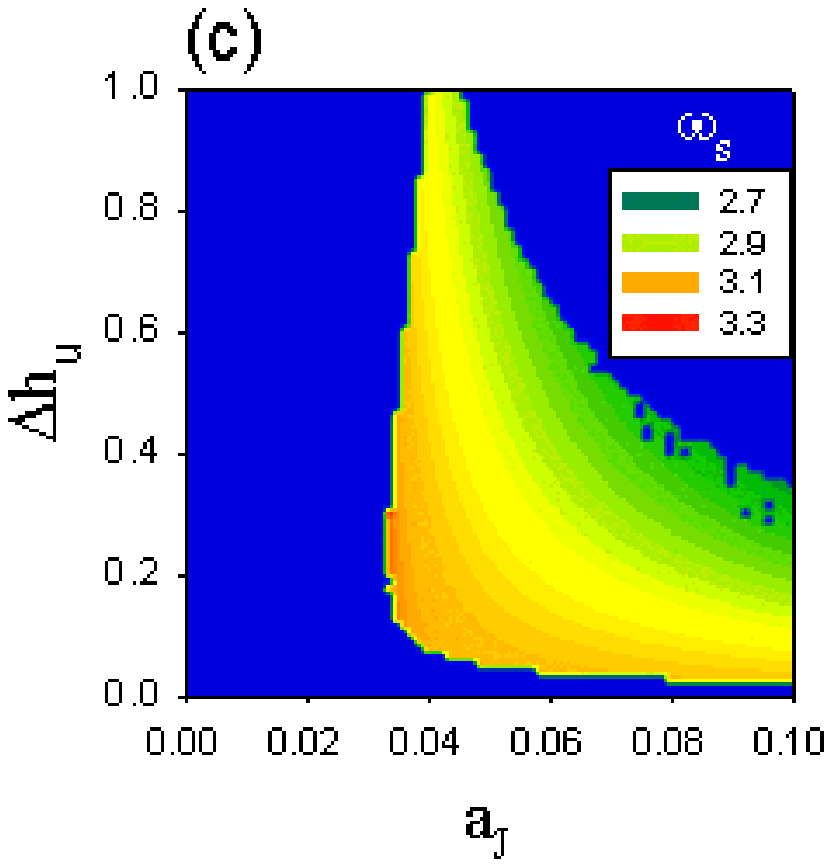}
\includegraphics[width=35mm]{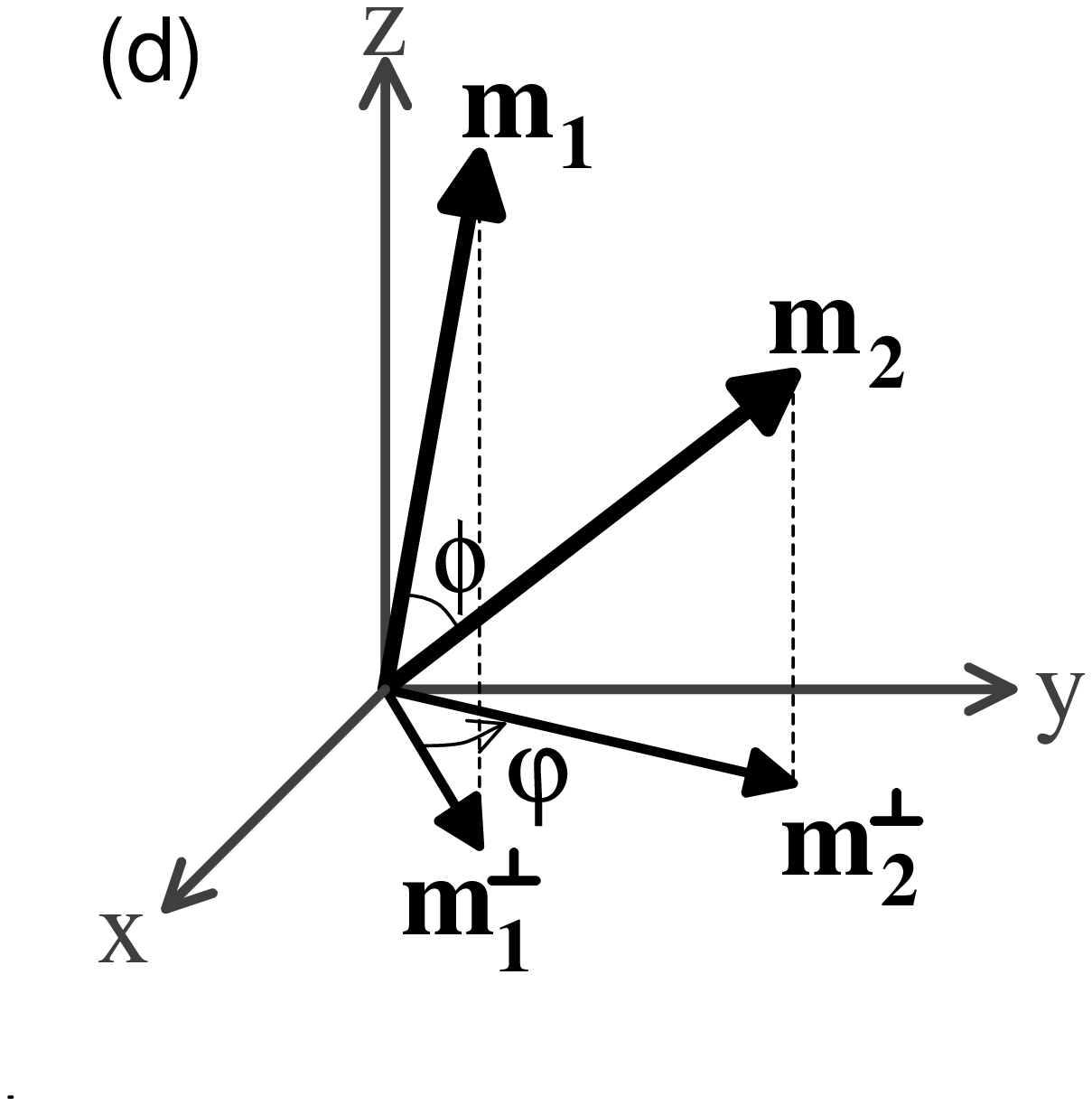}
\end{center}
\caption{(Color) 
         (a) Dependence of the phase shift $\varphi$ on $a_J$ and $\Delta h_\mathrm{u}$. 
         (b) Dependence of the angle between the two magnetizations $\phi$ 
         on $a_J$ and $\Delta h_\mathrm{u}$. 
         (c) Dependence of the precessional frequency $\omega_\mathrm{s}$ 
         in the synchronized precession on $a_J$ and $\Delta h_\mathrm{u}$. 
         (d) Definition of $\varphi$ and $\phi$.}
\label{fig:5}
\end{figure}

In the synchronized precession, the two magnetizations precess 
around the $z$ axis with the same period. 
Therefore, the phase shift $\varphi$ is locked. 
The magnitude of the locking phase shift depends on 
$a_J$ and $\Delta h_\mathrm{u}$ as is shown in Fig.~\ref{fig:5}(a). 
The numerical values in the legend of Fig.~\ref{fig:5}(a) 
represent the magnitude of $\varphi\slash \pi$. 
The phase shift becomes small as the magnitude of current increases. 

The angle between $\bm{m}_1$ and $\bm{m}_2$, 
$\phi=\arccos(\bm{m}_1 \cdot \bm{m}_2)$, also 
depends on $a_J$ and $\Delta h_\mathrm{u}$ in the synchronized precessional state. 
The dependence is shown in Fig.~\ref{fig:5}(b). 
The numerical values in the legend of Fig.~\ref{fig:5}(b) 
represent the magnitude of $\phi\slash \pi$. 
The angle between the two magnetization has the largest value 
in the precessional region 
where low $a_J$ and high $\Delta h_\mathrm{u}$. 

The precessional frequency $\omega_\mathrm{s}$ in the synchronized precession 
is tunable by current. 
In Fig.~\ref{fig:5}(c), we have shown the 
dependence of $\omega_\mathrm{s}$ on $a_J$ and $\Delta h_\mathrm{u}$. 
The legend in the figure represents the value of $\omega_\mathrm{s}$. 
It is found that $\omega_\mathrm{s}$ monotonously reduces 
as the magnitude of current increases.

Next, let us discuss the conditions for the excitation of the 
synchronized precession. 

In Fig.~\ref{fig:2}, we have shown the threshold for 
the current-driven synchronized precession. 
The threshold is denoted by the broken line L. 
We have found that L can be estimated by $a_J=\alpha h_\mathrm{c}^\mathrm{L}$ 
where $h_\mathrm{c}^\mathrm{L}$ is given by
\begin{equation}
h_\mathrm{c}^\mathrm{L}
=h_\mathrm{ext}^{(2)}+h_\mathrm{u}^{(2)}. 
\label{eq:threshold}
\end{equation}
Therefore, it is necessary that 
$a_J > \alpha h_\mathrm{c}^\mathrm{L}(=\alpha(3.0+\Delta h_\mathrm{u}))$ 
for the synchronized precession. 
It is noticed that the condition 
coincides with that for magnetization excitations in an asymmetric structure 
which can be obtained by a modified LLG equation; see eq.~(17) 
in ref. \citen{Sun2}. 
L corresponds to the threshold for magnetization excitations of $\bm{m}_2$ in 
a $\bm{m}_1$-pinned asymmetric trilayers. 
The reason that 
the current threshold L depends on 
the effective magnetic fields acting on $\bm{m}_2$ can be understood as follows; 
when the two magnetizations, $\bm{m}_1$ and $\bm{m}_2$, are initially 
almost parallel and 
a positive current ($a_J>0$) is applied, 
the direction of the spin-transfer torque is 
such that it induces the dynamics of $\bm{m}_2$ at first. 
On the other hand, 
when a negative current ($a_J<0$) is applied, 
the direction of the spin-transfer torque is 
such that it induces the dynamics of $\bm{m}_1$ at first.

It is noticed that 
the region P has spread in all ranges of $a_J$ 
for very small $\Delta h_\mathrm{u}$ in Fig.~\ref{fig:2}. 
Therefore, 
it is necessary that $\Delta h_\mathrm{u} \neq 0$ as well as that 
$a_J>\alpha h_\mathrm{c}^\mathrm{L}$ for the excitation of the synchronized precession. 
In other words, 
it is necessary for the synchronized precession 
that the effective magnetic fields acting on the two magnetizations 
are different. 
We generalize this necessary condition in \S \ref{sec:condition}.

Finally, we compare the results discussed above with 
the ones obtained by 
the macrospin model in the asymmetric structures. 
In the modified LLG model for the asymmetric structures, 
coherent magnetization precessions are possible only when 
$h_\mathrm{u}< h_\mathrm{ext} <h_\mathrm{u}+h_\mathrm{p}$ \cite{Gorley}. 
Here $h_\mathrm{ext}$, $h_\mathrm{u}(=1)$, and $h_\mathrm{p}$ are 
the dimensionless external magnetic field parallel to planes, 
in-plane uniaxial anisotropy field, and 
out-of-plane anisotropy field which act 
on the magnetization of the free layer, respectively. 
Accordingly, 
the out-of-plane anisotropy field $h_\mathrm{p}$ 
is important for current-induced magnetization precessions 
in the asymmetric structure. 
On the other hand, 
the synchronized magnetization precessions in the symmetric structure 
are possible even in the absence of $h_\mathrm{p}^{(i)}$. 
Actually, we have neglected the effect of $h_\mathrm{p}^{(i)}$ in this section.

\section{Magnetization Dynamics in Realistic Trilayers}
In this section, 
we discuss the magnetization dynamics 
including synchronized precessions in realistic trilayer structures. 
As a realistic setup, 
we use the set of parameters (ii): $\alpha=0.01$, $h_\mathrm{ext}^{(1)}=1.0$, 
$h_\mathrm{ext}^{(2)}=1.0+\Delta h_\mathrm{ext}$, 
$h_\mathrm{u}^{(1)}=h_\mathrm{u}^{(2)}=1.0$, and 
$h_\mathrm{p}^{(1)}=h_\mathrm{p}^{(2)}=10.0$.

\subsection{Phase diagram}
\begin{figure}
\begin{center}
\includegraphics[width=80mm]{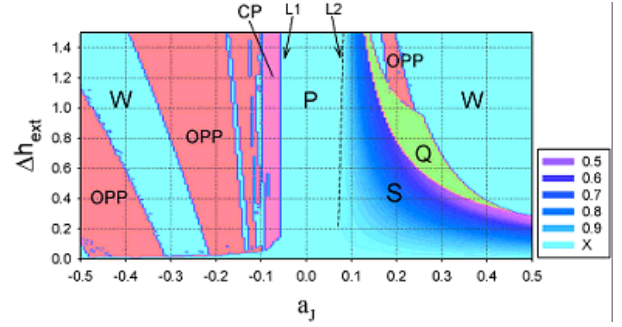}
\end{center}
\caption{(Color) Dynamical stability diagram 
for $a_J$ and $\Delta h_\mathrm{ext}$. S denotes synchronized precessions, 
Q quasi-periodic motions, 
OPP out-of-precessions, 
CP large-angle clamshell precessions, 
W chaotic behaviors, and 
P parallel configuration~\cite{note2}. 
The numerical values in the legend represent the values of 
$\langle m_{1z} \rangle$. 
X represents parallel configuration or chaotic behaviors. 
The broken lines, L1 and L2, represent the threshold for current-driven excitations.}
\label{fig:6}
\end{figure}

In Fig.~\ref{fig:6}, 
the dynamical stability diagram for $a_J$ and $\Delta h_\mathrm{ext}$ 
is shown. 
S, Q, OPP, CP, P, and W 
represent the region 
where synchronized precessions, quasi-periodic motions, 
out-of-plane precessions, clamshell precessions, parallel configuration, and 
chaotic behaviors occur, respectively. 
The numerical values in the legend represent the values of 
$\langle m_{1z} \rangle$ in the synchronized precession, 
where $\langle \cdots \rangle$ is the one cycle average. 
The broken lines, L1 and L2, are the threshold for current-driven excitations. 
L1 and L2 can be estimated by $a_J=-\alpha h_\mathrm{c}^\mathrm{L1}$ 
where 
\begin{equation}
h_\mathrm{c}^\mathrm{L1}
=h_\mathrm{ext}^{(1)}+h_\mathrm{u}^{(1)}+\frac{h_\mathrm{p}^{(1)}}{2}
=7.0,
\end{equation}
and $a_J=\alpha h_\mathrm{c}^\mathrm{L2}$ where 
\begin{equation}
h_\mathrm{c}^\mathrm{L2}=h_\mathrm{ext}^{(2)}+h_\mathrm{u}^{(2)}
+\frac{h_\mathrm{p}^{(2)}}{2}=7.0+\Delta h_\mathrm{ext}, 
\end{equation}
respectively. 
It is then necessary that $a_J<-\alpha h_\mathrm{c}^\mathrm{L1}$ or 
$a_J>\alpha h_\mathrm{c}^\mathrm{L2}$ for magnetization excitations.

Let us first discuss the positive current region ($a_J>0$) in Fig.~\ref{fig:6}. 

\begin{figure}
\begin{center}
\includegraphics[width=35mm]{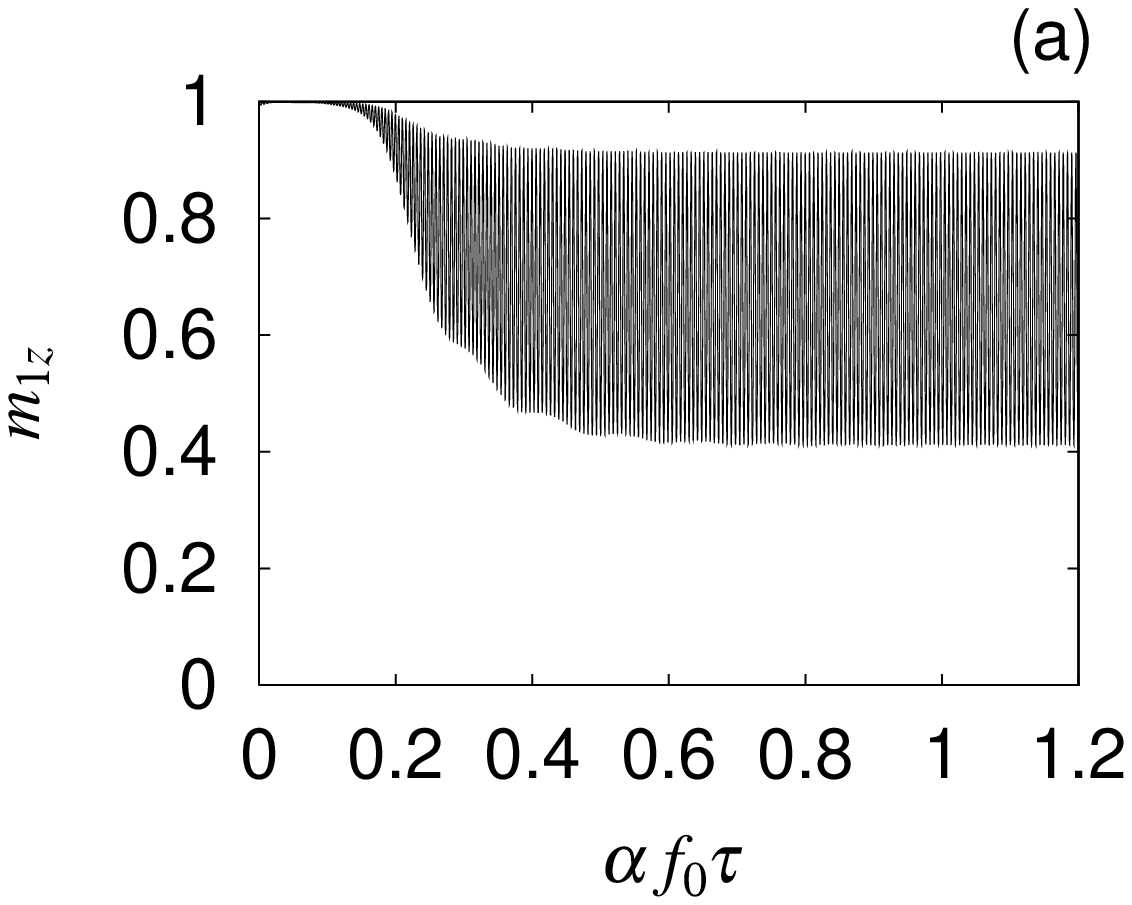}
\includegraphics[width=35mm]{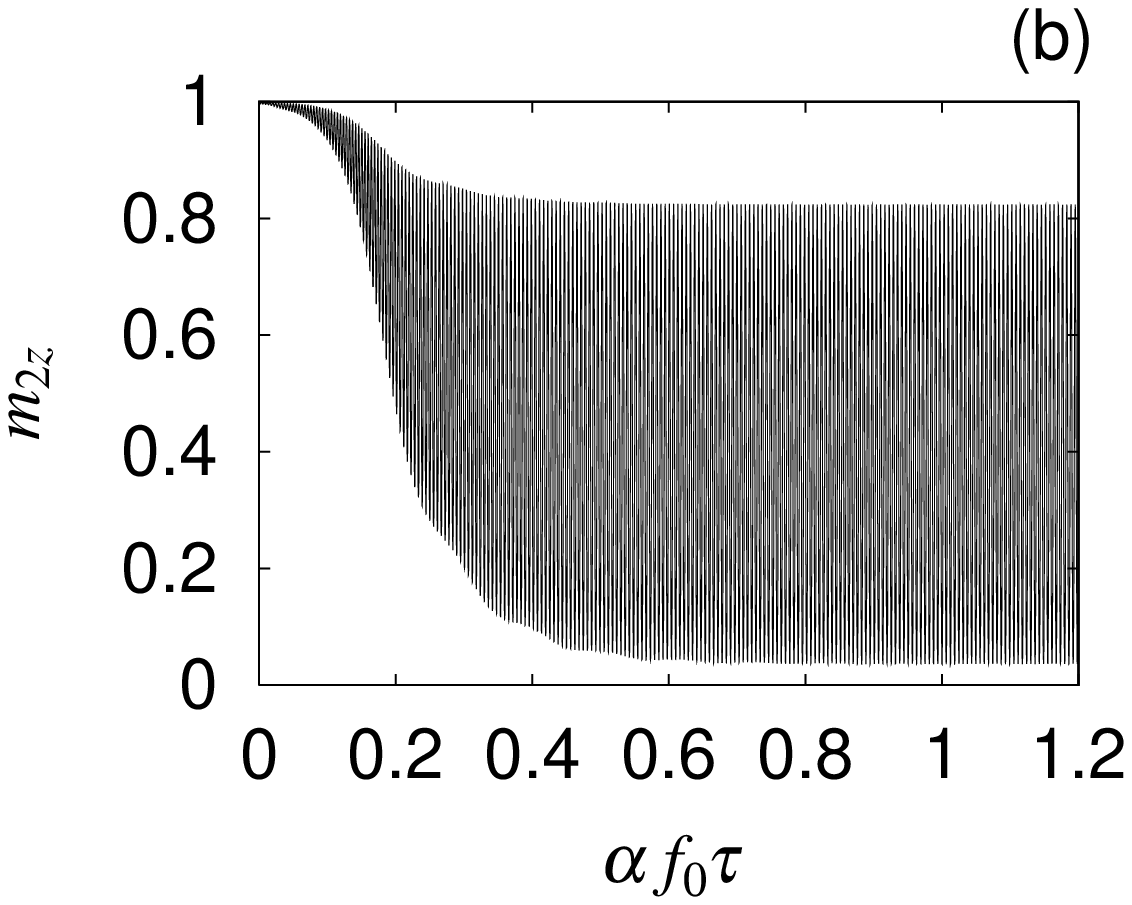}\\
\includegraphics[width=30mm]{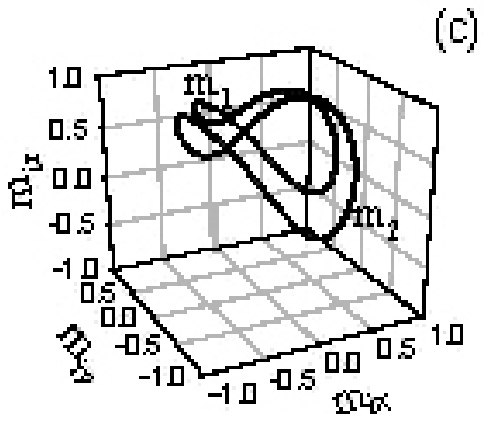}
\includegraphics[width=35mm]{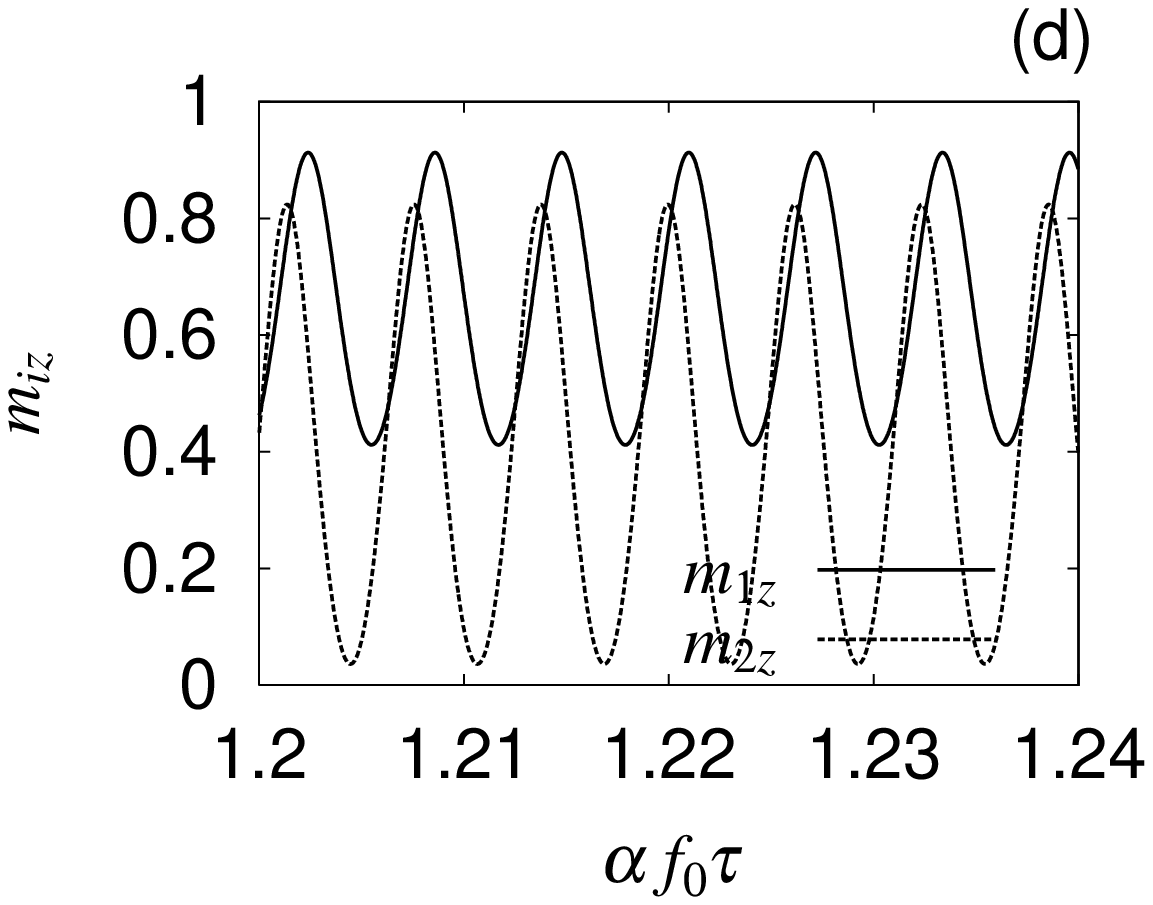}
\end{center}
\caption{Typical behaviors of $\bm{m}_1$ and $\bm{m}_2$ 
in the region S. 
The parameters are chosen as $a_J=0.2$ and $\Delta h_\mathrm{ext}=0.6$. 
The initial conditions are chosen as $m_{1z}(0)=0.999$, $m_{2z}(0)=0.998$, and 
$\varphi=\pi \slash 3$. 
(a) Time evolution of $m_{1z}(\tau)$. 
(b) Time evolution of $m_{2z}(\tau)$. 
(c) Bird's-eye view of the trajectories in the synchronized precession. 
(d) Time evolution of $m_{iz}(\tau)$ at intervals between 
$\alpha f_0 \tau=1.2$ and $1.24$.}
\label{fig:7}
\end{figure}

In the region S, 
both $\bm{m}_1$ and $\bm{m}_2$ reach the steady precessional states 
after transient behaviors as shown in Fig.~\ref{fig:7}(a) and \ref{fig:7}(b). 
In the steady state, 
the two magnetizations perform precessions around the $z$ axis with elliptic forms, 
whose trajectories are shown in Fig.~\ref{fig:7}(c). 
The elliptic forms of the precessions result from the effect of 
the effective demagnetizing field $h_\mathrm{p}^{(i)}$. 
As is shown in  Fig.~\ref{fig:7}(d), 
the precessional period of $\bm{m}_1$ and $\bm{m}_2$ are identical. 
That is, the phase shift between $\bm{m}_1$ and $\bm{m}_2$ is locked. 
Therefore, the synchronized precessions exist 
even in a realistic setup. 
The precessional amplitudes 
become large as $a_J$ and $\Delta h_\mathrm{ext}$ are increased. 
This tendency is the same as that discussed in the previous section.

\begin{figure}
\begin{center}
\includegraphics[width=32mm]{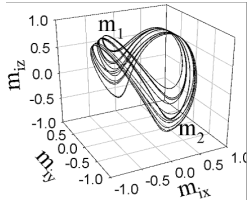}
\end{center}
\caption{Bird's-eye view of the trajectories in the region Q. 
The parameters are chosen as $a_J=0.3$ and $\Delta h_\mathrm{ext}=0.6$.}
\label{fig:8}
\end{figure}

\begin{figure}
\begin{center}
\includegraphics[width=45mm]{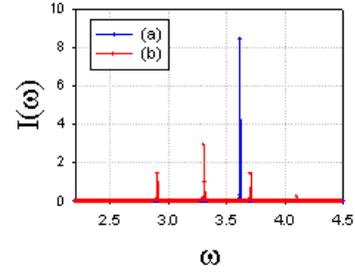}
\end{center}
\caption{(Color) 
(a) Typical power spectrum in the region S. 
The parameters are chosen as 
$(a_J,\Delta h_\mathrm{ext})=(0.24813,0.6)$. 
(b) Typical power spectrum in the region Q. 
The parameters are chosen as 
$(a_J,\Delta h_\mathrm{ext})=(0.24814,0.6)$.}
\label{fig:9}
\end{figure}

For larger $a_J$ or $\Delta h_\mathrm{ext}$, 
the phase locking become weak and 
quasi-periodic motions occur. 
The region where the quasi-periodic motions occur is denoted as Q in Fig.~\ref{fig:6}. 
The typical motions of $\bm{m}_1$ and $\bm{m}_2$ in the region Q are 
shown in Fig.~\ref{fig:8}. 
Both $\bm{m}_1$ and $\bm{m}_2$ perform precessions 
around the $z$ axis with several characteristic frequencies. 
The typical power spectrum for the quasi-periodic motions is 
shown in Fig.~\ref{fig:9}(b). 
$I(\omega) \equiv \lim_{\tau \to \infty} \frac{1}{\tau}
\langle m_{1x}^\ast(\omega) m_{1x}(\omega) \rangle$ is plotted, 
which is obtained by the Fourier transform of $m_{1x}(\tau)$ in the steady state. 
It is found that several characteristic frequencies coexist. 

The phase boundary between the region S and Q is very sharp. 
In Fig.~\ref{fig:9}(a) and \ref{fig:9}(b), we have shown the 
power spectrum for the steady-state motions at 
$(a_J,\Delta h_\mathrm{ext})=(0.24813,0.6)$ and 
$(a_J,\Delta h_\mathrm{ext})=(0.24814,0.6)$, respectively. 
It is determined by the very slight difference in $a_J$ 
which motions will occur between synchronized precessions 
and quasi-periodic motions. 
The change to quasi-periodic motions from synchronized precessions 
is not continuous for $a_J$, 
i.e., the change is drastic.

\begin{figure}
\begin{center}
\includegraphics[width=32mm]{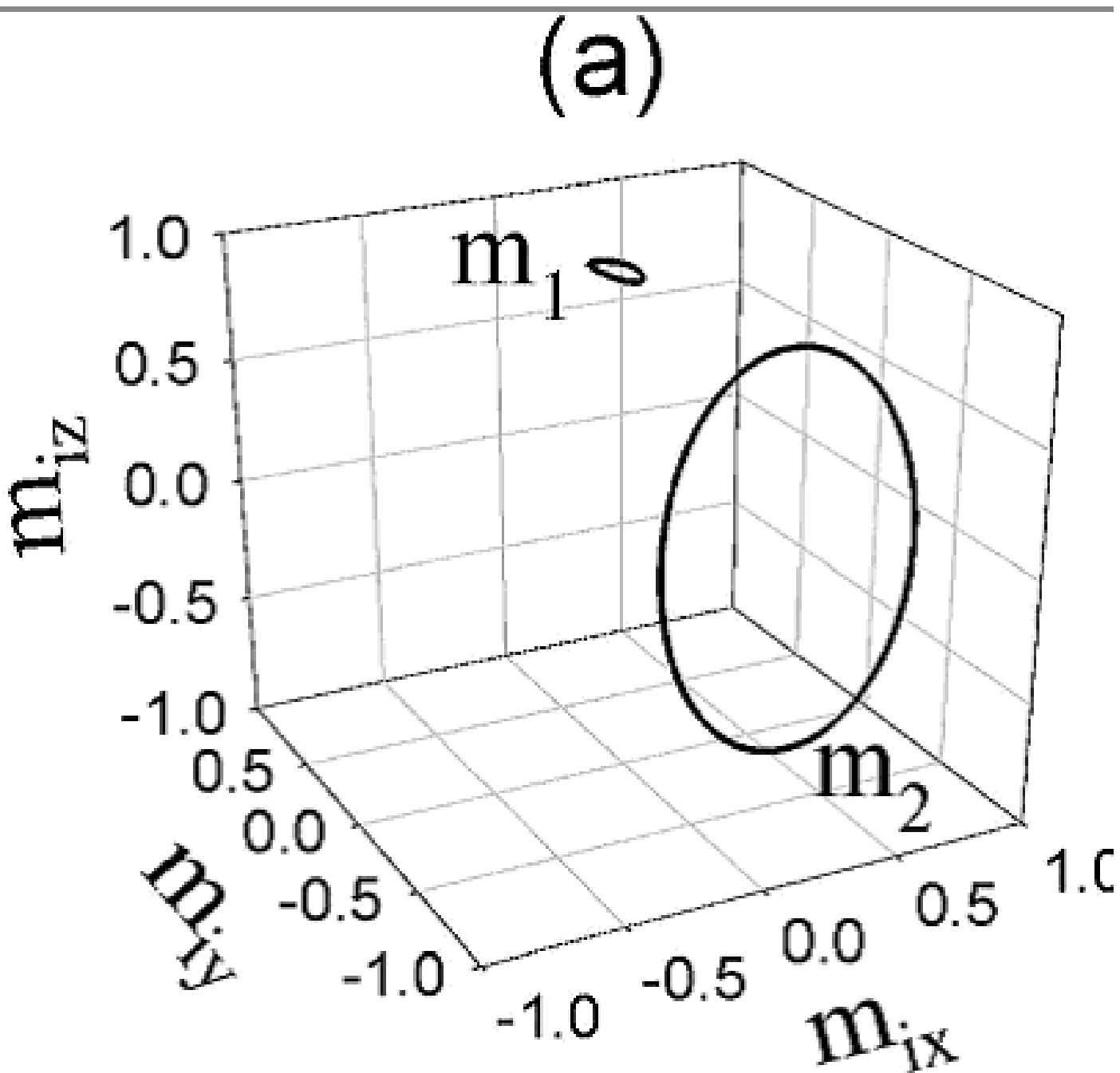}
\includegraphics[width=32mm]{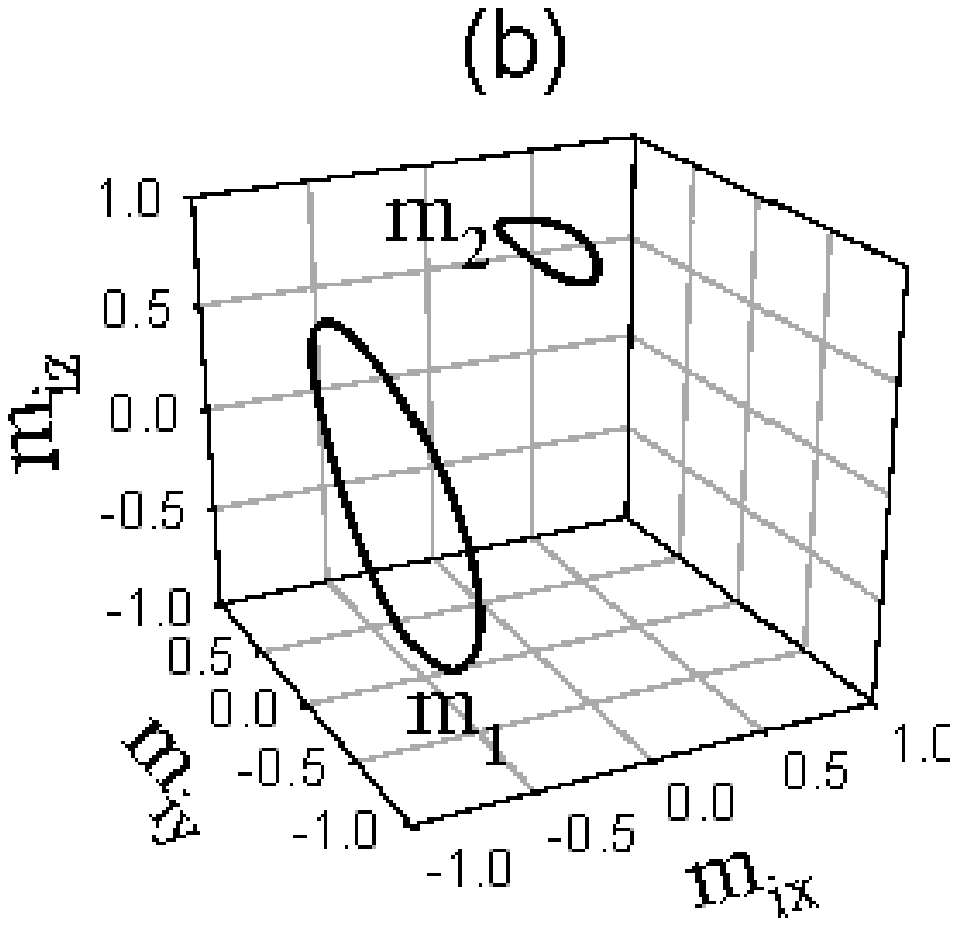}
\end{center}
\caption{Bird's-eye view of the trajectories in the region OPP. 
(a) Out-of-plane precession of $\bm{m}_2$. 
The parameters are chosen as $a_J=0.2(>0)$ and $\Delta h_\mathrm{ext}=1.2$. 
(b) Out-of-plane precession of $\bm{m}_1$. 
The parameters are chosen as $a_J=-0.45(<0)$ and $\Delta h_\mathrm{ext}=0.6$.}
\label{fig:10}
\end{figure}

In the region OPP, 
$\bm{m}_2$ performs the out-of-plane precession. 
On the other hand, $\bm{m}_1$ oscillates only around the vicinity of the 
positive $z$ direction. 
Those behaviors are shown in Fig.~\ref{fig:10}(a). 
As a matter of fact, 
like the magnetization dynamics in the region S, 
$\bm{m}_1$ and $\bm{m}_2$ also perform precessions with the same period 
in the region OPP. 
However, because 
one of them has a small precessional amplitude, 
the magnetization dynamics of the whole system is almost 
an out-of-plane precession.

\begin{figure}
\begin{center}
\includegraphics[width=35mm]{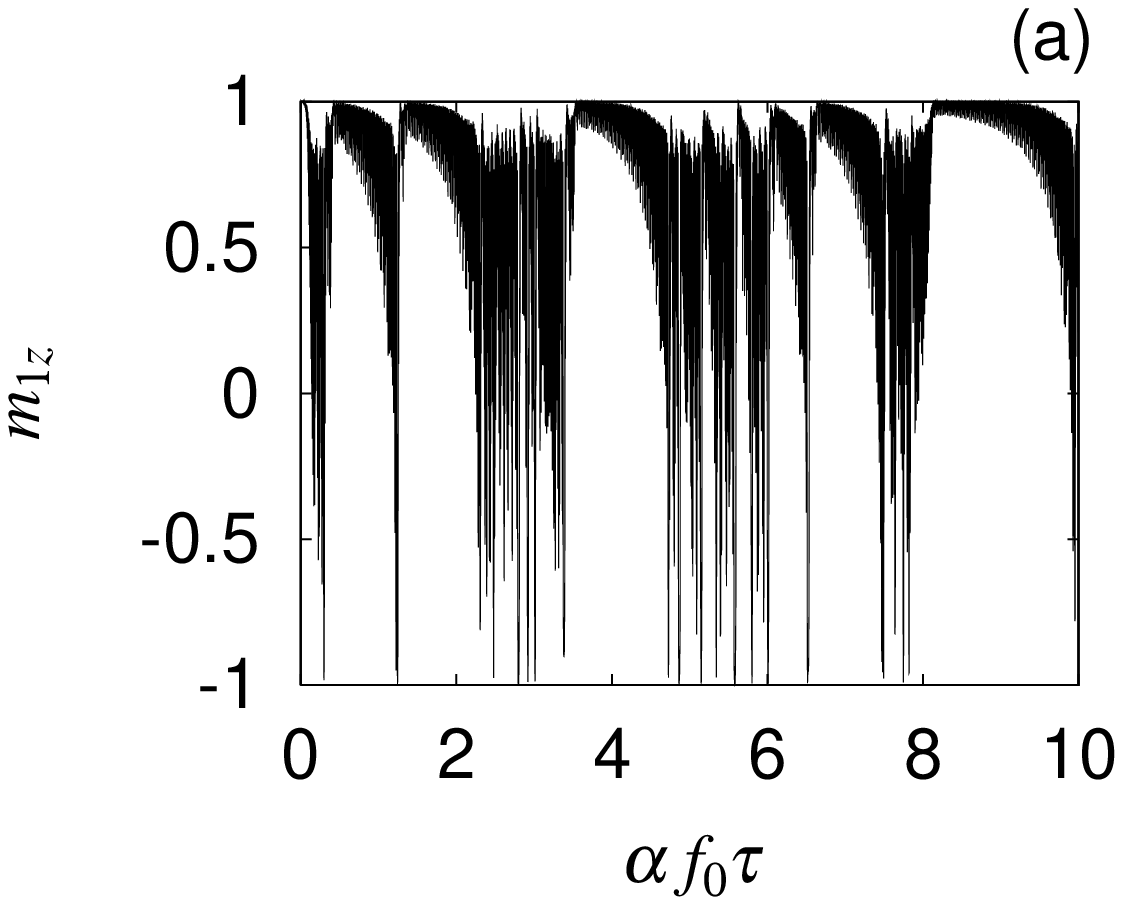}
\includegraphics[width=35mm]{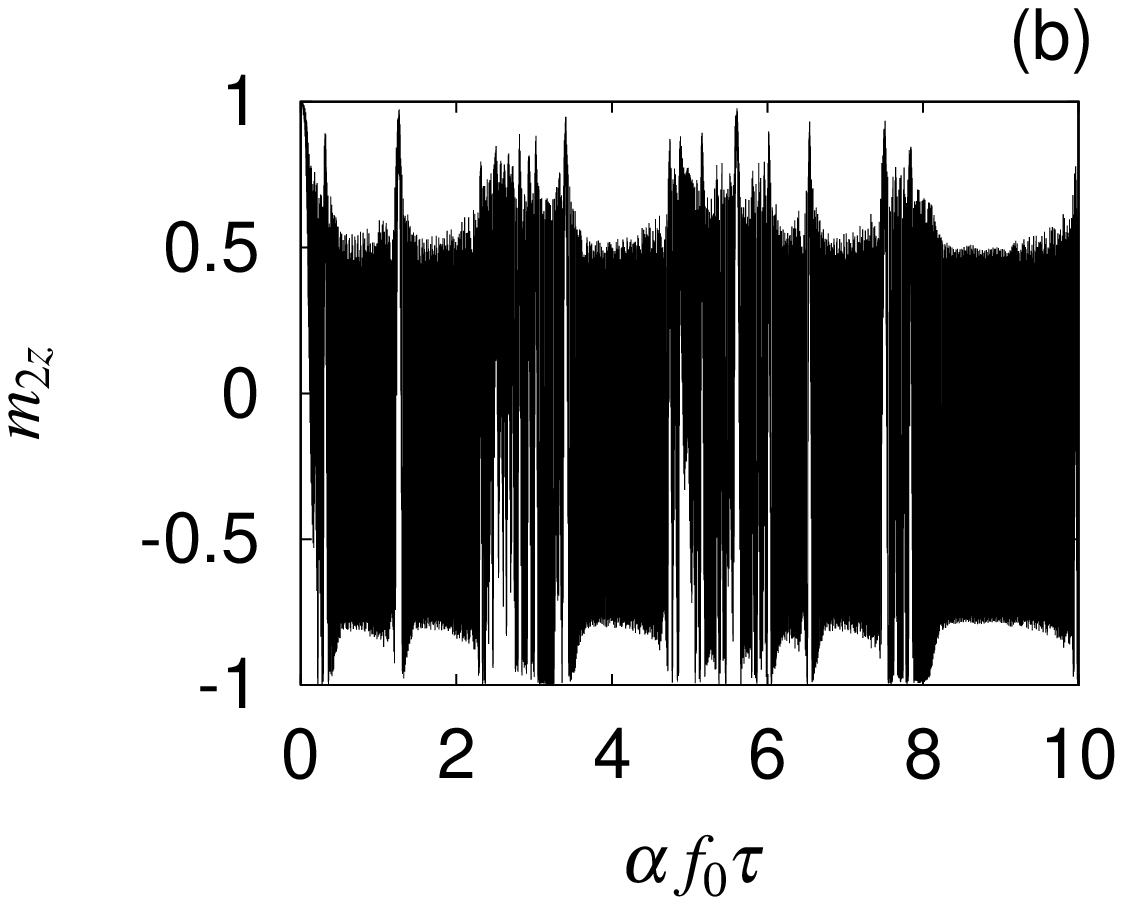}\\
\includegraphics[width=35mm]{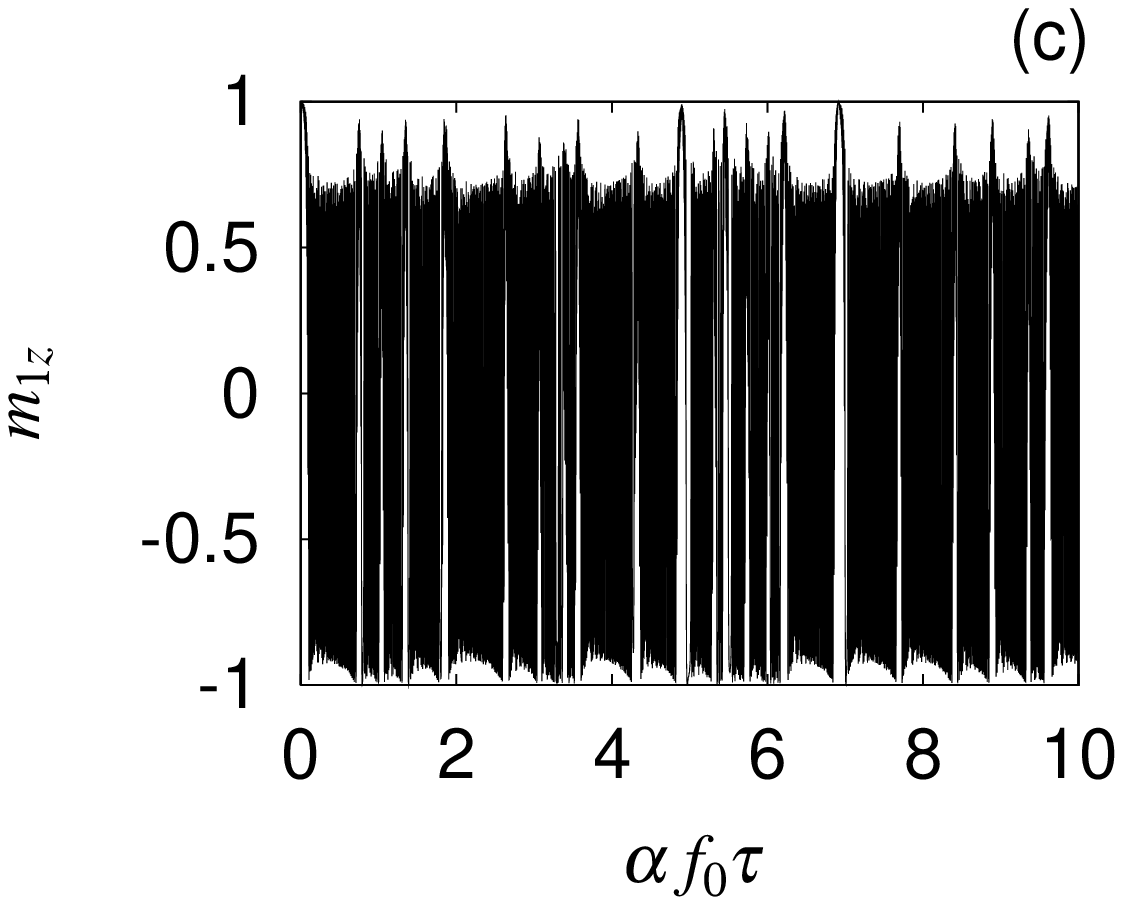}
\includegraphics[width=35mm]{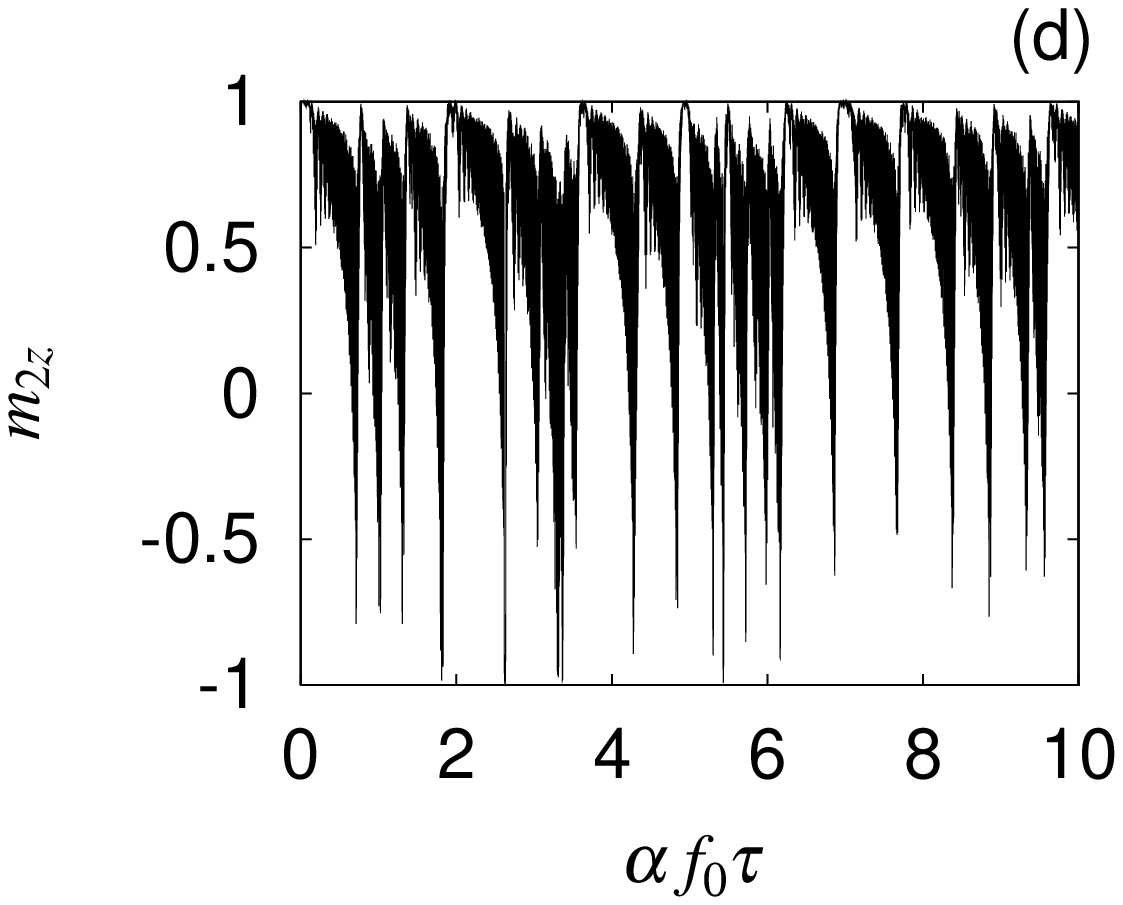}
\end{center}
\caption{Typical behaviors of $m_{iz}(\tau)$ in the region W. 
The initial conditions are chosen as $m_{1z}(0)=0.999$, $m_{2z}(0)=0.998$, and 
$\varphi=\pi \slash 3$. 
(a)-(b) Typical behaviors in the region W where $a_J>0$. 
The parameters are chosen as $a_J=0.4$ and $\Delta h_\mathrm{ext}=0.6$. 
(c)-(d) Typical behaviors in the region W where $a_J<0$. 
The parameters are chosen as $a_J=-0.35$ and $\Delta h_\mathrm{ext}=0.6$.}
\label{fig:11}
\end{figure}

For too large $a_J$, chaotic behaviors of $\bm{m}_1$ and $\bm{m}_2$ occur. 
The typical behaviors are shown in Fig.~\ref{fig:11}(a) and \ref{fig:11}(b) where 
time evolutions of $m_{1z}(\tau)$ and $m_{2z}(\tau)$ are plotted, respectively. 
$\bm{m}_1$ randomly changes its in-plane precessional amplitude. 
On the other hand, 
$\bm{m}_2$ hops between out-of-plane and in-plane trajectories irregularly.

Next, let us discuss the negative current region ($a_J<0$) 
in Fig.~\ref{fig:6}. 

In the region P, due to the smallness of current, 
any kind of magnetization dynamics is not excited. 
Just the static parallel configuration, $\bm{m}_1=\bm{m}_2=\hat{\bm{z}}$, 
occurs in the steady state. 

\begin{figure}
\begin{center}
\includegraphics[width=40mm]{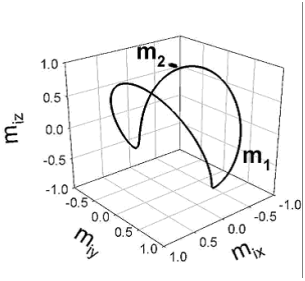}
\end{center}
\caption{Large-angle clamshell precession of $\bm{m}_1$ which occurs in the region CP. 
The parameters are chosen as $a_J=-0.085$ and $\Delta h_\mathrm{ext}=0.8$.}
\label{fig:12}
\end{figure}

For the current exceeding the threshold L1, 
$\bm{m}_1$ performs 
an in-plane large-angle clamshell precession and, 
on the other hand, 
$\bm{m}_2$ 
moves only in the vicinity of the $z$ axis. 
These typical behaviors are shown in Fig.~\ref{fig:12}. 
The behaviors become unstable at $a_J=-0.1$.

For $a_J<-0.1$, out-of-plane precessions of $\bm{m}_1$ or 
chaotic behaviors occur. 
In the region OPP, 
$\bm{m}_1$ performs the out-of-plane precession. 
On the other hand, $\bm{m}_2$ oscillates only around the vicinity of the 
positive $z$ direction. 
Those behaviors are shown in Fig.~\ref{fig:10}(b). 
In the region W, 
chaotic behaviors of $\bm{m}_1$ and $\bm{m}_2$ occur. 
The typical behaviors are shown in Fig.~\ref{fig:11}(c) and \ref{fig:11}(d) where 
time evolutions of $m_{1z}(\tau)$ and $m_{2z}(\tau)$ are plotted, respectively. 
$\bm{m}_1$ hops between out-of-plane and in-plane trajectory irregularly.
$\bm{m}_2$ randomly changes its in-plane precessional amplitude. 
It is noted that according to the sign of current,
the behavior of $\bm{m}_1$ and $\bm{m}_2$ is contrastive 
as can be recognized by comparing Fig.~\ref{fig:10}(a) with \ref{fig:10}(b) 
and comparing 
Fig.~\ref{fig:11}(a)-(b) with \ref{fig:11}(c)-(d). 
These behaviors in the region OPP and W reflect that 
the positive current region is the $\bm{m}_2$-excited region 
and the negative current region the $\bm{m}_1$-excited region. 

In the negative current region ($a_J<0$), 
the magnetization dynamics of the whole system is almost determined 
by $\bm{m}_1$ because $\bm{m}_2$ oscillates with very small amplitudes 
as mentioned above. 
Moreover, the obtained coherent behaviors in $a_J<0$ are 
similar to those obtained by a macrospin model in an asymmetric structure: 
in-plane precessions and out-of-plane precession \cite{Kiselev}. 
Accordingly, roughly speaking, the magnetization dynamics 
is very similar to those in an asymmetric structure, 
especially its in-plane high magnetic field region; 
see Fig.~3 in ref. \citen{Kiselev}, for example. 
The difference is that there exists chaotic behaviors in our model.

\subsection{Microwave frequency}
\begin{figure}
\begin{center}
\includegraphics[width=50mm]{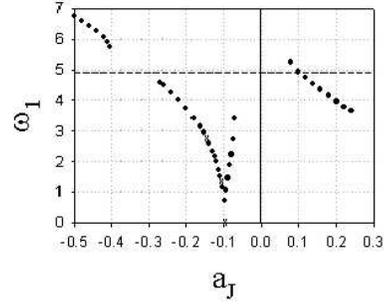}
\end{center}
\caption{Dependence of the precessional frequency $\omega_1$ on $a_J$ 
         when $\Delta h_\mathrm{ext}=0.6$. 
         The broken line represents $\omega_0 \simeq 4.9$ determined by 
         eq.~(\ref{eq:w_0}).}
\label{fig:13}
\end{figure}

Figure~\ref{fig:13} shows 
the dependence of $\omega_1$ on current $a_J$. 
$\omega_1$ is the precessional frequency of $\bm{m}_1$. 
$\Delta h_\mathrm{ext}$ is fixed as $\Delta h_\mathrm{ext}=0.6$. 
The dependence of $\omega_1$ on $a_J$ reflects the behaviors shown in Fig.~\ref{fig:6} 
with $\Delta h_\mathrm{ext}=0.6$. 

For the negative current regime ($a_J<0$), 
we have obtained the dependence of microwave frequency on current 
similar to the one which can be obtained in the asymmetric structure~\cite{Kiselev}. 
In the in-plane precessional region, 
the precessional frequency 
$\omega_1$ decreases as the magnitude of current $|a_J|$ increases. 
In the out-of-plane precessional region, 
$\omega_1$ increases as $|a_J|$ increases. 
$\omega_1$ can not be defined in the region W 
because the characteristic frequency does not exist in chaotic behaviors. 
Therefore, in Fig.~\ref{fig:13}, we have not plotted the data in the region W or 
have labeled several points by cross marks. 

For the positive current regime ($a_J>0$), 
we have plotted the dependence of $\omega_1$ on $a_J$ 
in the synchronized precession. 
The synchronized precessional frequency 
monotonously decreases as the magnitude of current is increased. 
This tendency is the same as that discussed in the previous section; 
see Fig.~\ref{fig:5}(c).

\subsection{Conditions for the excitation of synchronized precessions}\label{sec:condition}
As can be found from Fig.~\ref{fig:6}, 
it is necessary that $\Delta h_\mathrm{ext} \neq 0$ 
as well as $a_J>\alpha h_\mathrm{c}^\mathrm{L2}$ 
for the excitation of synchronized precessions. 
That is, the difference between the effective fields acting on two magnetizations 
is needed. 
This condition can be generalized; 
it is necessary that there exists the deviation between 
the intrinsic frequency of F1 and F2. 
The intrinsic frequency of $\bm{m}_i$ is given by 
$\omega_0^{(i)}=
\sqrt{(h_\mathrm{ext}^{(i)}+h_\mathrm{u}^{(i)})
(h_\mathrm{ext}^{(i)}+h_\mathrm{u}^{(i)}+h_\mathrm{p}^{(i)}})$. 
Accordingly, the above condition can be written as 
$\omega_0^{(1)} \neq \omega_0^{(2)}$. 
We have checked the condition using several sets of parameters. 
The condition that $\omega_0^{(1)} \neq \omega_0^{(2)}$ 
can be understood in the following way. 
If $\omega_0^\mathrm{(1)}=\omega_0^\mathrm{(2)}$, 
$\bm{M}_1$ and $\bm{M}_2$ tend to 
arrange their motions and become parallel to each other. 
As the result, the spin-transfer torque $\bm{\Gamma}_i$ can not work. 
This condition is always applicable when $\bm{M}_1$ and $\bm{M}_2$ 
have their equilibrium points in the same direction 
in the absence of current. 

It is noted that synchronized precessions can not exist 
when the deviation of intrinsic frequencies, 
$\Delta \omega=\omega_0^{(2)}-\omega_0^{(1)}$, is too large; $\Delta \omega \gg 0$. 
$\Delta \omega$ originates in 
the deviation of the effective magnetic fields 
such as $\Delta h_\mathrm{u}$ or $\Delta h_\mathrm{ext}$. 
As it is recognized in Fig.~\ref{fig:2} or \ref{fig:6}, 
$\Delta \omega$ must be in a limited range, 
otherwise chaotic behaviors occur. 
One of the reason is that the phase locking of two magnetization precessions 
is impossible when $\Delta \omega$ is too large.

\subsection{Comparison with macrospin models in an asymmetric structure} 
We compare the behaviors obtained by eqs.~(\ref{eq:LLG}) 
with the ones obtained by the macrospin model in an asymmetric structure. 
It is known that two types of coherent magnetization motions 
of a free layer can be obtained by an modified LLG equation in an asymmetric structure: 
in-plane precessions and out-of-plane precessions. 
Chaotic behaviors like telegraph noise can not be obtained 
by such an macrospin model 
because the degree of freedom of the magnetization is only two. 
Accordingly, the ``W'' phase 
experimentally observed by Kiselev {\it et al.} \cite{Kiselev} 
can not be obtained. 
In our model, however, 
chaotic behaviors can be obtained by eqs.~(\ref{eq:LLG}) 
since the degree of freedom of the system is four. 
The region W appearing the negative current regime 
in Fig.~\ref{fig:6} may corresponds to 
the ``W'' phase discussed in ref.~\citen{Kiselev}.

\section{Conclusions}
Current-induced magnetization dynamics 
in a symmetric trilayer structure consisting of two ferromagnetic free layers 
and a nonmagnetic spacer is examined. 
We have treated the two free layers as a monodomain ferromagnet 
and calculated the magnetization dynamics by modified LLG equations. 
We have found that various behaviors of the two magnetizations 
arise depending on the applied current and the deviation of 
effective magnetic fields acting on them. 
Especially, there exists the synchronized precessions of two magnetizations 
among the various behaviors. 
For the excitation of the synchronization, 
current must exceed the threshold whose magnitude is the same as 
that for magnetization excitations in an asymmetric structure. 
It is also necessary that 
the intrinsic frequencies of the two magnetizations 
are different; $\omega_0^{(1)} \neq \omega_0^{(2)}$. 
Moreover, the deviation of frequencies, $\Delta \omega=\omega_0^{(2)}-\omega_0^{(1)}$, 
must be in a limited range for the phase locking of two magnetization precessions. 
Utilizing the synchronization in a symmetric trilayer 
may be one of the possible ways to raise the microwave emission power 
of the spin-transfer oscillators \cite{Kaka,Mancoff,Grollier2}.

\section*{Acknowledgment}
The authors would like to thank the staff of 
Frontier Research Laboratory in Toshiba Research \& Development center  
for their daily help.


\begin{thebibliography}{99}
\bibitem{note} Since $\bm{m}_i \cdot \dot{\bm{m}_i}=0$ from eqs.~(\ref{eq:LLG}), 
it always holds that $|\bm{m_i}|=1$.
\bibitem{note2} The boundary between the region Q and OPP is very rough. 
              At some points in Q near OPP, out-of-plane precessions also 
              occur depending on initial conditions.
\end{thebibliography}
\end{document}